\documentclass[journal]{IEEEtran}
\usepackage{graphicx}
\usepackage{multirow}
\usepackage{hyperref}
\usepackage{url}
\usepackage{placeins}
\usepackage{subcaption}
\usepackage{algorithm}
\usepackage{algpseudocode}

\title{A Modular, End-to-End Next-Generation Network Testbed: Towards a Fully Automated Network Management Platform}

\author{Ali~Chouman, Dimitrios Michael~Manias, and
        Abdallah~Shami \thanks{Ali Chouman, Dimitrios Michael Manias, and Abdallah Shami are with the Department of Electrical and Computer Engineering at Western University e-mail: \{achouman, dmanias3, Abdallah.Shami\}@uwo.ca}}

\begin{document}

\maketitle

\begin{abstract}
Experimentation in practical, end-to-end (E2E) next-generation networks deployments is
becoming increasingly prevalent and significant in the realm of modern networking and wireless communications research. The prevalence of fifth-generation technology (5G) testbeds and the emergence of developing networks systems, for the purposes of research and testing, focus on the capabilities and features of analytics, intelligence, and automated management using novel testbed designs and architectures, ranging from simple simulations and setups to complex networking systems; however, with the ever-demanding application requirements for modern and future networks, 5G-and-beyond (denoted as 5G+) testbed experimentation can be useful in assessing the creation of large-scale network infrastructures that are capable of supporting E2E virtualized mobile network services. To this end, this paper presents a functional, modular E2E 5G+ system, complete with the integration of a Radio Access Network (RAN) and handling the connection of User Equipment (UE) in real-world scenarios. As well, this paper assesses and evaluates the effectiveness of emulating full network functionalities and capabilities, including a complete description of user-plane data, from UE registrations to communications sequences, and leads to the presentation of a future outlook in powering new experimentation for 6G and next-generation networks. 
\end{abstract}

\maketitle

\section{Introduction}

Modern telecommunication networks face a remarkable challenge in supporting a large number of user devices and connections with stringent quality-of-service (QoS) requirements. To address such a challenge, data-driven approaches and solutions have been employed to improve future designing and decision-making in modern and future networks. In the fifth-generation of technology standards, machine learning (ML) and artificial intelligence (AI) techniques power network intelligence engines that leverage real-time data collected within the network, howsoever dynamic the network environment may be; however, research in wireless communications has emphasized the need for end-to-end (E2E) network systems, or full-scale emulation 5G+ testbeds, that have practical use in real network deployments \cite{hoeschele20225g}. These system prototypes, or fully functional modular systems, require ease of access and integration with intensive applications leveraging the core network, such as the 5G Core (5GC) in the 5G standard. As such, there is a need for a fully automated network management platform, as a testbed or full system, to act as a basis for basic network operations inside a wireless network deployment. In addition, the platform will enable novel network data analytics solutions and advanced intelligence for emerging network use cases, such as anomaly detection and performance optimization in the realm of network management.

The E2E 5G network requires multiple components, or modules, that span over all domains of a wireless network; the design of E2E networks implies that some layer of management and orchestration, such as ETSI Network Function Virtualisation Management and Orchestration (ETSI NFV MANO) is essential to ensuring all working components meet defined service-level agreements (SLAs) \cite{pateromichelakis2019end}. Furthermore, communication service providers (CSPs) who leverage these E2E network systems are concerned with Quality-of-Service (QoS) and Quality-of-Experience (QoE) postulates that will depend on the system configuration and available network resources to satisfy such demands \cite{li20215growth}. As such, an E2E network is confined to the all User Equipment (UE) devices, or UEs, connecting through a Radio Access Network (RAN) to a core network (\textit{e.g.,} the 5G Core) in order to access an application or service. The nature of E2E networks is, then, the realization of the E2E service chain that CSPs must utilize to ensure proper service operations for devices in a wireless network \cite{alliance20175g, HawiloJSAC, MoubayedTMC}.

The traditional architectures of wireless communications systems cannot sustain the strenuous demands of new-and-coming applications in future networks. Mobile Network Operators (MNOs), who already own and operate the physical infrastructure of such networks, are assessing and exploring the microservices operated by core networks in the design of a service-based architecture (SBA) \cite{yang2016service}. A higher level of network management, and corresponding automation, will be required in future networks as such architectures will be able to support automated decision-making within the network without the need for human intervention \cite{liyanage2022survey}. These network management platforms are greatly useful to Mobile Virtual Network Operators (MVNOs), who provide their virtual services atop the existing physical infrastructure that they have leased: automated network management allows MVNOs to optimize their services' performances and maintenance without hindering existing MNO deployments \cite{ashraf2022zero}.

The contributions of this work are summarized as follows:
\begin{itemize}
    \item The assessment of the architectural composition of an E2E network management system as a testbed, and detailed insights into all its modular components, compliance with the Third Generation Partnership Project (3GPP) specifications and protocols, and functional/non-functional requirements.
    \item The evaluation of modern developing network management platforms in the context of E2E network systems.
    \item A 3GPP-compliant, modular, E2E next-generation network testbed-as-a-system that spans the entire core network, RAN, UEs, and applications/services for service providers.
    \item An implemented Network Data Analytics Function (NWDAF) within the network for in-house advanced analytics and automated intelligence.
    \item The discussion on transitioning to 6G and future networking oriented towards network management platforms for MNOs, CSPs, and business consumers of next-generation applications and solutions. 
\end{itemize}

\section{Designs and Implementations of E2E Networks in Literature}

The following section discusses the need for synthesizing E2E networks and network management platforms for E2E network systems, and the motivation for realizing an E2E network system towards supporting fully automated network management. Thereafter, an assessment of previous designs and implementations in literature is provided to support the aforementioned motivation.\par

Given that 5G+ networks are expected to service a multitude of devices with stringent network requirements, experimentation and field trials in the 5G context become a necessity for academic work \cite{gringoli2018performance}. Research in 5G applications and development must be on par with prominent industry projects and as a result, 5G testbeds can give researchers an open platform for testing a variety of use cases that are beneficial to business consumers. Therefore, realizing a closed 5G testbed is not sufficient for testing service operations; rather, an E2E network system that connects user devices to a data/application network through an access network and core network consolidates all these components for ease of access and management. An E2E 5G+ network system is a real deployment of the core and access networks in contrast to a testbed that emulates these functionalities for the sole purpose of testing simulations.\par

The following highlights some of the key implementations and advancements derived from prototypes leveraging these two technologies. When building a modular E2E network system, there are two main open-source 5G core implementations: \textit{Open5GS} and \textit{free5GC}. For reference, a comparison between these two implementations, and other leading open-source 5G Core solutions, is provided in Table \ref{experiment_params}. The 5GC can be further split into the Control Plane (CP) and the User Plane (UP) as per the Control and User Plane Separation Principle (CUPS). The 5GC is designed to support services and data connectivity with Network Functions (NFs), enabling deployment using enabling technologies such as Network Function Virtualisation (NFV) and Software Defined Networking (SDN). The need for these novel techniques is increasing due to the multitude of microservices offered by the service-based architecture of the 5G network \cite{tsai_lin_tanaka_2021}. Another comparison between open-source 5G core implementations, such as \textit{Open5GS} and \textit{free5GC}, is important for understanding the state-of-the-art. The authors compare the three implementations based on the attributes: infrastructure, base projects, maturity, community, and management. Based on the author’s analysis, \textit{Open5GS} was selected as the best software for their implementation based on community activity and a built-in orchestration framework. Since the release of this work, however, \textit{Open5GS} and \textit{free5GC} have become the most prevalent open-source 5G core implementations.\par

\textit{Open5GS}, which is a C-language open-source implementation of 5GC, is also Release-16 compliant, or in accordance with the 3GPP release specifications \cite{lee}. The 5GC NFs provided by \textit{Open5GS} include the following: the Access and Mobility Management Function (AMF), in addition to the Network Repository Function (NRF), Session Management Function (SMF), Authentication Server Function (AUSF), Unified Data Management (UDM), Unified Data Repository (UDR), Policy Control Function (PCF), Network Slice Selection Function (NSSF), Binding Support Function (BSF), and User Plane Function (UPF). A single gNB connects to the fully operational 5GC and a single host uses multiple network interfaces to emulate different UE devices interacting with the RAN and the UPF. All NFs are run as Linux executable programs in each Virtual Machine (VM). Barrachina-Munoz \textit{et al.} \cite{barrachina2022cloud} develop a cloud-native 5G platform with this technology and other monitoring platforms. Their implementation leverages \textit{Open5GS} and \textit{Prometheus}-based monitoring deployed using \textit{Kubernetes} for containerization. Furthermore, they integrate Amarisoft Callbox, a commercial 3GPP-compliant gNB, to complete their E2E prototype. The authors go on to present two use cases stemming from their prototype: UPF re-selection and UE mobility. Alcalde Cespedes \cite{alcalde2022log} proposes log-based monitoring and automatic anomaly mitigation in 5GC. \par

\textit{free5GC} is used in a large span of academic works for its collective featuring of a more modern 5G standalone architecture \cite{foukas2018experience}. The authors have developed the Model Training Logical Function (MTLF) and the Analytics Logic Function (AnLF), as per the 3GPP specification for 5G Core operations \cite{3gpp.23.288}. In their work, the authors introduce the most prominent 5G core open-source projects, OAI, \textit{Open5GS}, and \textit{free5GC}, with the justification for choosing \textit{free5GC} being that other projects are still in development. The authors list extensive experimental studies of the implemented NWDAF as an avenue for future work. Chu \textit{et al.} \cite{jain2022l25gc} propose restructuring the 5G core network to reduce the latency in the control plane. The authors leverage \textit{free5GC} to implement their proposed solution. Based on the presented results, the proposed work improves the performance of the 5GC by reducing latency in the control plane compared to the basic \textit{free5GC} deployment. Liu \textit{et al.} \cite{9973025} conduct a performance analysis on \textit{free5GC}'s forwarding in private and public clouds. The authors leverage Amazon Web Services (AWS) for the public cloud and their own deployment and configuration for the private cloud. They explore using different acceleration technologies and realistic traffic models during the performance testing. The authors compare and contrast the deployment performance on the private and public clouds and comment on the current limitations of each. Lee and Shin \cite{lee2022federated} consider using federated learning in private 5G networks. The authors use \textit{free5GC} to configure a distributed NWDAF. The presented scenario considers multiple factories acting as federated nodes contributing to the federated learning process. The presented demo considers an image recognition application with the central model being stored in the NWDAF, which also acts as the federated learning aggregation agent.\par

NextEPC was tested in the realized E2E network system in a similar fashion to related works because associated literature discuss the implementations of an open-source 5G solution with this software. In particular, it features traffic differentiation, as packets can be forwarded in many different ways across network connection points; however, this poses a major architectural revision to the NWDAF implementation as it would need to trace multiple interfaces for any given NF-NF interaction \cite{chen2020realization}. A proposed mechanism using NextEPC and OpenStack for network slicing was considered in the same NFVI architectural fashion as investigated in this paper.\par

OpenAirInterface (OAI) is a popularly used open-source software implementation of the 4G mobile network, and was considered in the system design. Disregarding the control plane differences in the 4G EPC with relation to the 5G Core, Open Air Interface (OAI) features a built-in emulation capability for scalable execution in real environments. The main advantage over solutions like \textit{Open5GS} and \textit{free5GC} is that OAI offers two physical layer emulation modes (PHY) that improve the real RAN interactions between the UE and gNB. This is in contrast to \textit{Open5GS}, for example, which is commonly used with \textit{UERANSIM} to emulate the radio physical layer using the \textit{UERANSIM} Link Protocol. With regards to commercial deployment, OAI is known to be expensive and not as flexible as the other solutions \cite{nikaein2014openairinterface}, so the proposed E2E network system uses \textit{Open5GS} for experimentation.\par

Other solutions considers the development of an E2E network, as well as a novel implementation and integration of the NWDAF with open-source frameworks. The following works are derived from the insights gathered by the NWDAF. Firstly, an initial analysis of the signalling traffic and the BSF/NRF interaction was conducted \cite{chouman2022}. A subsequent work considered the characterization of NF-NF interactions based on unsupervised learning \cite{maniasgc2022}. Furthermore, a model drift detection and adaptation framework for core networks is considered in \cite{manias2022model}.\par

A developed 5G+ testbed can also consider open-source RAN simulators, including my5G-RANTester, an Amarisoft gNB and UEs, and a monitoring framework. Additionally, the author has developed an automation framework that detects unregistered users and adds them to the network database. Sylla \textit{et al.} \cite{sylla2022emu5gnet} proposes an open-source emulator for 5G SDNs. Their proposed framework integrates \textit{UERANSIM} and \textit{Open5GS} with emulated WiFi networks and a Virtual Infrastructure Manager to provide NFV MANO functionalities. The authors present a vehicular use case where various vehicular clients connect to both 5G and WiFi access points. A data processing pipeline is created where the vehicle-generated data is transmitted to the closest data center located in the simulated environment. Apostolakis \textit{et al.} \cite{apostolakis2022design} develop an E2E emulation of a 5G mobile network using various open-source technologies. The authors leverage \textit{srsRAN} to emulate UEs and the access network, \textit{Open5GS} is used for core simulation, and GNU Radio Companion is used for wireless channel emulation. A \textit{Prometheus} deployment is used to ensure ubiquitous network monitoring. The authors present two experimental scenarios considering dynamic orchestration and UE mobility. Choudhari \textit{et al.} \cite{choudhari2022deployment} consider the deployment of 5G core private networks. The authors use \textit{Open5GS} and \textit{UERANSIM}. The authors outline the steps required to integrate these two open-source implementations in terms of configuring the Access and Mobility Management Function (AMF), setting up the gNB, and accessing the gNB GUI. Furthermore, the authors discuss the creation of the NAT network containing the virtual machines hosting the software and the configuration such that they can access the public internet. Ungureanu and Vladeanu \cite{ungureanu2022leveraging} propose the containerized deployment of \textit{Open5GS} in the public cloud with a \textit{Kubernetes}-native packet manager. The authors argue that previous deployments only consider the virtualization capabilities of \textit{Open5GS}, whereas their work also considers the service-based model. This is achieved through the analysis of service-level communication between NFs. The authors use \textit{UERANSIM} as a UE and RAN emulator. The results presented consider AMF and NRF service response throughput. Tan \textit{et al.} \cite{tan2020reliable} propose an intelligent routing mechanism for 5G core networks. The authors' solution leverages \textit{free5GC} to develop their experimental testbed. In their evaluation, the authors created a topology with 11 UPFs with simulated bandwidth, latency, and loss to explore the mechanics of their algorithm. The authors compare their solution with the existing round-robin, and dynamic load balancing solutions and show that their proposed method leads to lower loss, greater throughput, and significantly lower latency.\par

The RAN and UE emulation involve the main components of the E2E network system that emulate an actual RAN technology and can either simulate the physical layer of radio interactions with the UE or connect real UE devices to a base station.

The modularity of the E2E 5G+ system allows for a large variety of emerging RAN solutions, or technologies that encompass the access network so that user devices can access applications through the core network. Traditionally, this is a mobile access network that devices connect to using the available radio resources, but it can also include non-terrestrial access points, such as satellites \cite{kavehmadavani2023intelligent}.

\textit{UERANSIM}, an open-source state-of-the-art 5G UE and RAN implementation, is a popular component that is typically found in conjunction with \textit{Open5GS}, to complete full operation of the 5G Core with connected devices \cite{aligungr}. Its main feature is providing basic functionality for gNB stations and UE devices as processes within the 5G system. As well, it is designed for the 5G standalone architecture which is important when studying specific 5G scenarios and not LTE-5G scenarios. The main disadvantage of this technology is that it does not provide the complete physical layer in 5G NR radio interactions. The radio interface is partially implemented and is simulated over the User Datagram Protocol (UDP).

An important uSIM implementation with \textit{free5GC} is proposed in \cite{ahmadi20195g} as another solution considering the automation platform it discusses. The authors consider the HTTP streaming application with their user technology and the zero-touch aspect of the solution is enabled through the NSSF and AMF core network functions. The authors use NFV Management and Orchestration as another architecture proposed and outlined for the 5G Core. The authors discuss the Open Network Automation Platform (ONAP), which leverages Enhanced Control, Orchestration, Management \& Policy (ECOMP) as well as the Open Orchestrator (Open-O) to enable complete network function lifecycle management. The authors go on to discuss the procedure of creating a network slice through the ONAP orchestrator. The RAN and UE can be analyzed by a scalability and performance analysis when interacting with the 5G Core. The authors base their methodology on Performance Evaluation Process Algebra \cite{gilmore1994pepa} and evaluate the  scalability in terms of concurrent users, NF multiplicity, and QoS. The authors perform an extensive analysis through two illustrative case studies considering session establishment and V2X user registration.

The topic of improved user equipment and RAN emulation has gained significant attention in recent years. Sattar and Matrawy \cite{sattar2019optimal} explore optimal RAN emulation in 5G core networks. The authors consider both the physical isolation of gNBs across network slices, as well as guaranteed E2E delay. The authors extend the formulation proposed by Dietrich \textit{et al.}  \cite{dietrich2017network}, which considered the placement of gNBs in the 5G Core. The presented results explore the effect of the level of slice isolation on CPU and bandwidth utilization, as well as the percentage of accepted requests. The authors complement their work by highlighting the security benefits of slice isolation in terms of mitigating DDoS attacks on the core network \cite{sattar2019towards}. \par

\begin{table*}[!htbp]
\centering
\caption{Features and Limitations of Prominent Open-Source 5G Core Solutions}
\label{experiment_params}
\begin{tabular}{|p{2.3cm}|p{6.7cm}|p{6.5cm}|}
\hline
\multicolumn{1}{|c|}{\textbf{5G Solution}} & \multicolumn{1}{|c|}{\textbf{Features}} & \multicolumn{1}{|c|}{\textbf{Limitations}}  \\ 
\hline
\textbf{\textit{Open5GS}}            & •	Release-16 compliant             &     •	Lack of roaming and emergency call capabilities                                \\
                                       & •	Supports USIM cards and multiple PDU sessions for UEs               & •	No Interworking with EPC                                   \\
                                       & •	IPv4 and IPv6 support                 & •	No NB-IoT or OCS/OFCS                                  \\
                                       & •	Handover sequences                  & •	No eMBMS                                   \\
                                       & •	CSFB (Circuit Switched Fall Back) and SMSoS (SMS Over SGs)                 &                                    \\
                                       & •	Supports VoLTE(Voice over LTE) and VoNR(Voice over NR)                  &                                    \\
                                       
\hline
\textbf{\textit{free5GC}}                 & •	Virtualizable and scalable, and runs on containers, virtual machines following Control and User Plane Separation (CUPS) design principle               &        •	Centralized architecture in the form of a cluster                             \\
                                       & •	Leverages multi-core processing components for optimal performance                       & •	AMF and SMF functionalities are not fully segregated                                   \\
                                     
\hline
\textbf{\textit{NextEPC}}                 & •	Proven to support large number of IoT devices in practice (approximately 50 000 registered UE devices)              &         •	Performance and capacity in real-time have yet to be assessed                                \\
                                       & •	3GPP compliant                       & •	No active deployments in other open-source solutions                                 \\
                                       & •	Supports 10 Gbps rates                    &                             \\
                         
\hline
\textbf{\textit{OpenAirInterface}}                 & •	Readily connectable to VNFs and NFV frameworks              &                    •	Limited by centralization and virtualization of gNBs/eNBs when implementing ORAN technologies                 \\
                                       & •	Supports MME load-balancing                      &                                    \\
                                       & •	Overloading control                   &                                    \\
                                       \hline

\end{tabular}
\end{table*}

The following presents an analysis of the functional and non-functional requirements of an E2E 5G+ system and the assessment of such requirements in other 5G prototypes and testbeds in literature. This includes the comparison and contrast of open-source software implementation used to realize a 5G testbed, and the evaluation of the validity and novelty of these testbeds.

A functional NWDAF implementation must develop and incorporate a working 5G core, and valuable insights must be drawn from preliminary analyses of generated and collected data. The conglomeration of data sources into an analytics function repository can demonstrate the uses of the NWDAF in applications focused on intelligent network management, and assess the current limitations of 5G networks as a future outlook for developments and technologies in 6G networks.

There is a consensus among related works in emerging E2E prototypes and testbeds that the key functionality of the 5G Core, or Beyond 5G core networks, is the interaction with the UE and the 
RAN. Neither of the two components use protocols 
which work in conjunction with the service-based architecture. The procedures of individual protocols, which are not 
handled alone by reference points across NF interfaces, must therefore be mapped onto the SBA 
in the form of APIs \cite{corici2022organic}.

\section{A Realized Modular E2E 5G Network}

\subsection{Design and Architecture}

The following section demonstrates the practical 5G network deployment as a functional, modular, E2E system. In detail, the system design and architecture is displayed and elaborated on, along with descriptions on user plane data analysis, performance metrics \& monitoring, network system security, and core/user data privacy.

\begin{figure*}[!htbp]
\centerline{\includegraphics[width=1.3\textwidth, height=0.8\textwidth, angle=90]{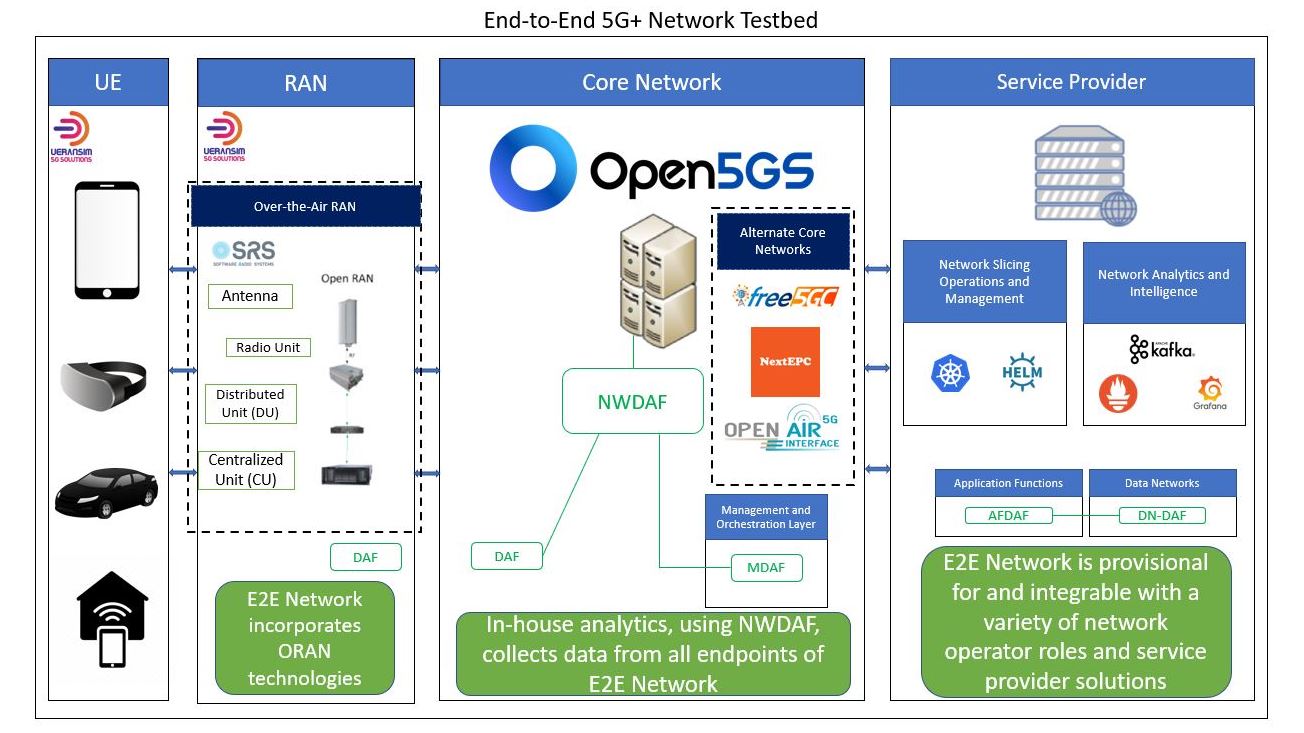}}
\caption{Diagram of E2E 5G+ System}
\label{systemproto}
\end{figure*}

Figure \ref{systemproto} shows the E2E 5G+ system with the UE, RAN, and 5G Core being the foundational components of this system, and the last column elucidates the capabilities of the system to interface with communications and business operator roles depending on their purposes and use cases for 5G testing. The following sub-sections decompose the E2E 5G+ system discussed in this paper into its main components, with the aim of illustrating the motivation and need for prototyping local 5G developments and solutions. In particular, an E2E 5G+ system comprises of the core network (\textit{e.g.,} 5G Core), the Radio Access Network, and the User Equipment, and a detailed background is provided for the architectural components and the technologies used to implement them in academic literature. The fourth component, titled "Service Provider", illustrates how CSPs employ their own solutions and other third-party solutions in service management and operations that can be integrated into the E2E 5G+ system.

Each module or component of the realized E2E 5G+ system represents an important integration that must complete the system before MNOs, CSPs, and interested third parties may leverage the whole E2E network. Using virtualization, the core network is represented as a single server hosting multiple virtual machines that themselves host an entity of the E2E system. For example, a core NF or a UE device each reside in a separate virtual machine. Containerization, however, is a more viable alternative in the management and orchestration of the full E2E system. \textit{Kubernetes} is used for implementing each core network component as a Containerized Network Function (CNF) and deploying the 5G Core onto the cluster. The other modules of the E2E system can reside on the host machine as well. For automated network management, many platforms already require the use of a \textit{Kubernetes} cluster over basic virtualization or bare-metal infrastructure.  Using \textit{Kubernetes} to orchestrate the containerized core network is challenging because IP addresses are assigned to each pod after scheduling, and this poses a challenge for the NWDAF instance to keep track of all NFs if more than one instance of an NF is introduced in the network. For example, it is not possible to provide static IP allocation to a UPF pod, hence the change in UPF configuration to ensure normal operation.

\begin{figure*}[!htbp]
\centerline{\includegraphics[width=2\columnwidth]{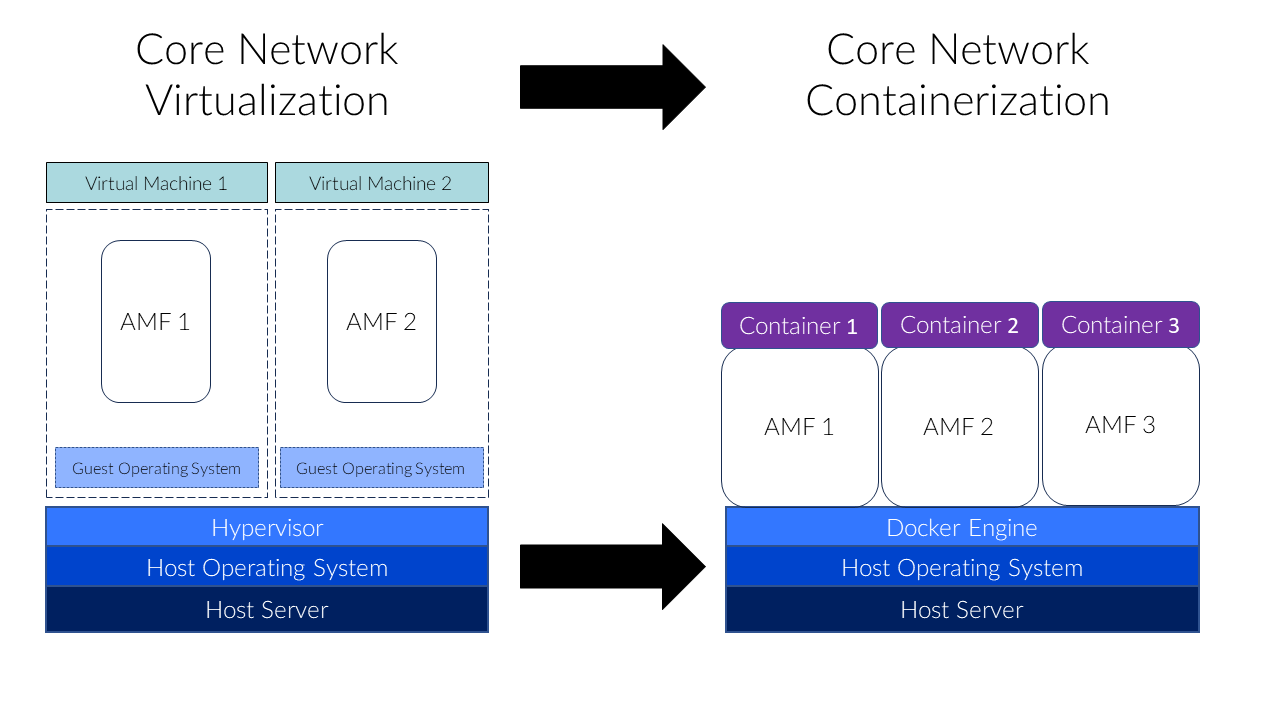}}
\caption{Evolution of Core Network Management and Orchestration}
\label{container}
\end{figure*}

Figure \ref{container} illustrates the evolution of network function management and orchestration in next-generation networks. Specifically, E2E testbeds experiment with Virtualised Network Functions (VNFs), where the host OS relies on the hypervisor to instantiate individual VMs for each core NF. For example, Figure \ref{container} shows multiple AMF instances in separate VMs, complete with a full guest OS. In early E2E system prototyping, the host server running the core network uses Windows Server as its host OS. The hypervisor manages multiple VMs with Ubuntu 22.04 as the guest OS, which allows for open-source core solutions to run the core NFs as Linux processes. Over bare metal infrastructure, virtualization is important for segregating core functionalities in different locations across the network; however, running the core NFs in individual containers allows the NFs to share host OS resources and avoid the instantiation overhead of allocating resources to a separate physical location in the network. Figure \ref{container} shows the Docker Engine that is employed by \textit{Kubernetes} to manage different core NFs as containers in its own private network. When each container is configured to communicate with each other, it presents a more efficient alternative to orchestrating NFs as CNFs. As well, containerization offers quick NF deployment and scaling, which are essential to automated network management in dynamic environments \cite{osmani2021multi}. 

The E2E 5G+ system consists of an open-source C++ implementation of the 5G Core standalone architecture using \textit{Open5GS} \cite{lee} and another open-source C-language implementation of the 5G NR RAN and UE devices using \textit{UERANSIM} \cite{aligungr}. The 5G Core also introduces the NWDAF as an operating NF within the Core, using an analytics engine built and integrated within the 5GC standalone architecture. The primary objective of the NWDAF, as defined by the 3GPP, is to aid in decision-making within the network based on prescribing and predicting network data. 

\subsection{User Data Analysis and E2E Validation}

User plane data analysis and protocol study are vital to understanding UE interactions with the core network as well as evaluating aspects of wireless communications as functional and architectural requirements of the core network. For the 5G Core, it is imperative that emerging network deployments be ready to support the strict reliability and latency requirements of certain UE use case classes, such as Ultra-Reliable Low-Latency Communications (uRLLC). Therefore, the realized E2E 5G+ system is structured to support uRLLC requirements, such as redundant data transfers, by architectural configuration. The 3GPP outlined three supporting requirements for uRLLC in the technical specfication, TS 23.501 \cite{homma_miyasaka_matsushima_voyer}.

\begin{figure}[!htbp]
\begin{subfigure}{.5\textwidth}
  \centering
  \includegraphics[width=1\linewidth]{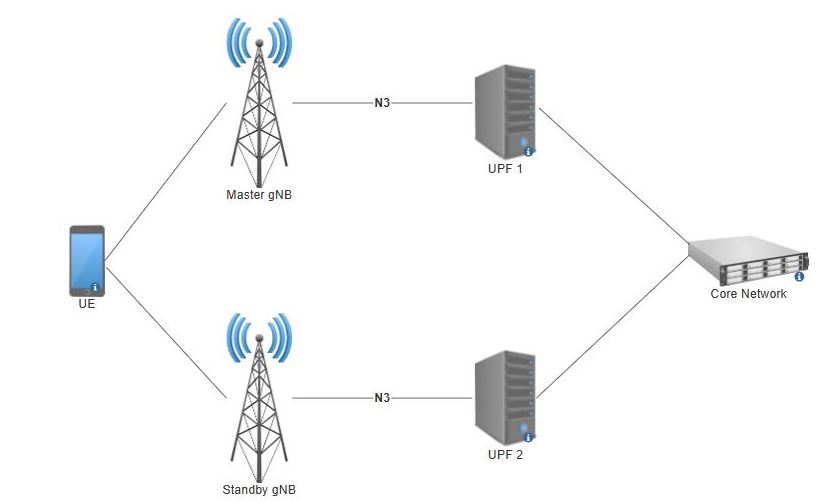}
  \caption{Dual Connectivity Supported by RAN and Core Network}
  \label{fig:sfig1}
\end{subfigure}
\begin{subfigure}{.5\textwidth}
  \centering
  \includegraphics[width=1\linewidth]{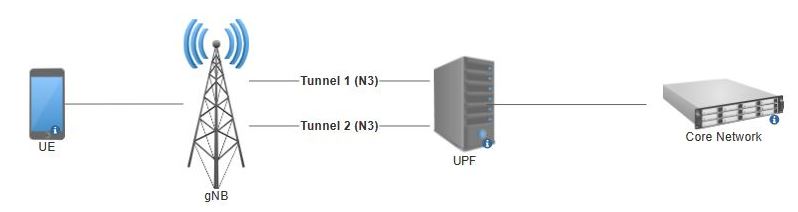}
  \caption{Packet Replication via UPF Tunneling}
  \label{fig:sfig2}
\end{subfigure}
\begin{subfigure}{.5\textwidth}
  \centering
  \includegraphics[width=1\linewidth]{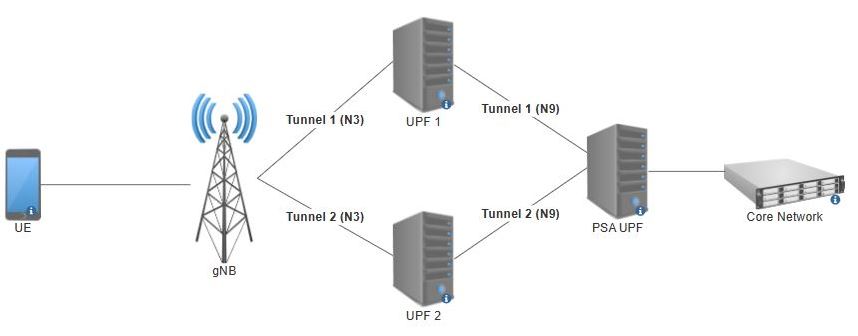}
  \caption{Mobile User Plane Evolution with PSA UPF Model}
  \label{fig:sfig3}
\end{subfigure}
\caption{User Plane Protocol Requirements Supporting uRLLC Devices for E2E 5G+ System}
\label{fig:fig}
\end{figure}

Figure \ref{fig:fig} highlights the architectural requirements realized in the RAN and core network of the E2E 5G+ system in order to support uRLLC devices. Figure \ref{fig:sfig1} shows the first requirement, which entails redundant transmission paths as PDU sessions that replicate packets between the UE and the data network. Through the standby gNB and the second UPF instance, which are both orchestrated and maintained by the E2E system management, a second path for redundant data transfer in uRLLC is established. Figure \ref{fig:sfig2} illustrates the next requirement as packet replication through tunneling, in contrast to the first requirement as a wholly separate path. Packets between the UE and a UPF instance use GPRS Tunneling Protocol (GTP) encapsulation, and two separate N3 tunnels are created to replicate packets both on the uplink and downlink between the RAN and the UPF instance. Finally, Figure \ref{fig:sfig3} displays the third requirement as the evolution of edge networking using a PDU Session Anchor (PSA) UPF instance. Individual packet tunnels for each UPF instance will route their traffic through the common PSA UPF as a method of replication and UPF GTP packets can forward the same sequence number in the IP packet.

\begin{figure}[!htbp]
\centerline{\includegraphics[width=1\columnwidth]{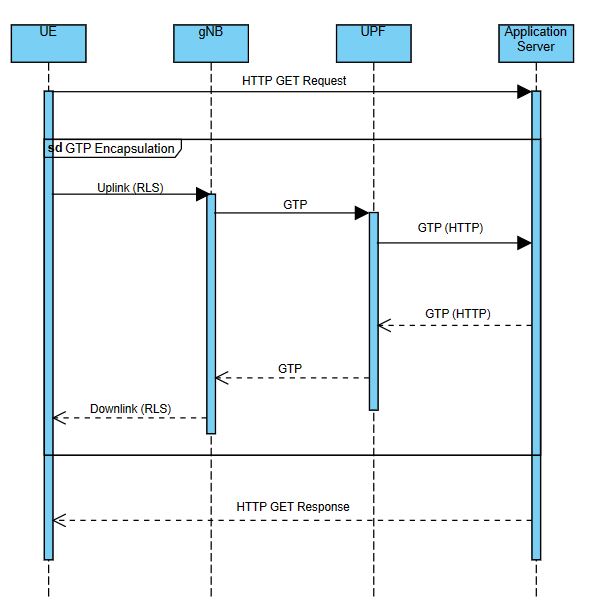}}
\caption{Sequence Diagram of UE Application Requests through 5G Core}
\label{sequence}
\end{figure}

Figure \ref{sequence} is a sequence diagram that demonstrates the proper procedure of a UE device requesting from an HTTP application server through the 5G Core. An identical sequence can be identified in the dataset provided by \cite{chouman2022} using the 5G system in early prototyping, and was made publicly available \cite{western-oc2-lab}. HTTP requests from the UE device have their destination address as the server they wish to reach; however, the 5G Core routes these packets from the RAN side to the UPF through GTP encapsulation. The E2E system shows the example request as uplink traffic to the nearest gNB instance using the Radio Link Simulation (RLS) protocol provided by \textit{UERANSIM} for the physical layer. Thereafter, the gNB communicates to the UPF using GTP tunneling, with the UPF as the destination, as the UPF will be responsible for routing packets to the final destination through the 5G Core. When the response is received from the application server, the UPF sets the new destination address to the gNB, so that the UE device can download the response through the gNB as downlink radio traffic.

To ensure full E2E operations in the entire system requires a validation of successful registrations in the core and RAN, in addition to verifying that certain networking protocols enable all service operations across the E2E system. For the core network, validating the registration of each NF and their corresponding services from the perspective of the SBI is essential to ensuring proper E2E network operations. 

\begin{figure*} [!htbp]
\begin{subfigure}{2\columnwidth}
  \centering
  \includegraphics[width=\columnwidth]{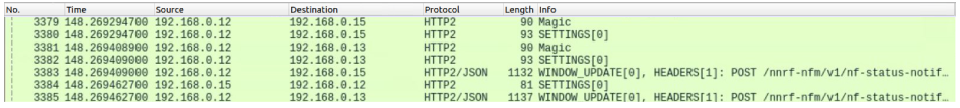}
  \caption{HTTP/2 NF Services over SBI}
  \label{mfig:mfig1}
\end{subfigure}\\ \\
\begin{subfigure}{2\columnwidth}
  \centering
  \includegraphics[width=\columnwidth]{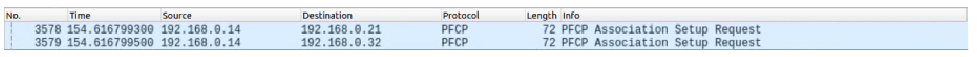}
  \caption{SMF-UPF Association}
  \label{mfig:mfig2}
\end{subfigure} \\ \\
\begin{subfigure}{2\columnwidth}
  \centering
  \includegraphics[width=\columnwidth]{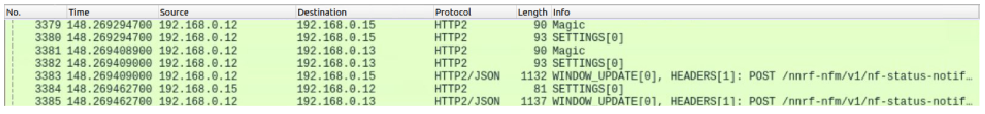}
  \caption{AMF-gNB Association During gNB Registration}
  \label{mfig:mfig3}
\end{subfigure} \\ \\
\begin{subfigure}{2\columnwidth}
  \centering
  \includegraphics[width=\columnwidth]{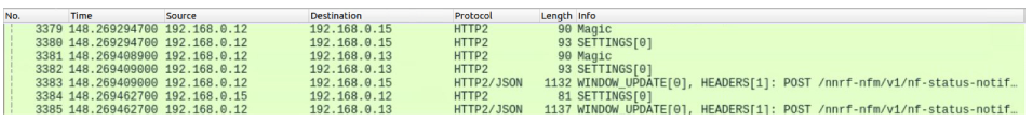}
  \caption{NF Heartbeats}
  \label{mfig:mfig4}
\end{subfigure} \\ \\
\begin{subfigure}{2\columnwidth}
  \centering
  \includegraphics[width=\columnwidth]{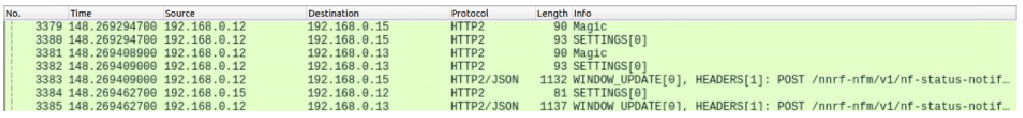}
  \caption{Full UE Registration}
  \label{mfig:mfig5}
\end{subfigure} \\ \\
\begin{subfigure}{2\columnwidth}
  \centering
  \includegraphics[width=\columnwidth]{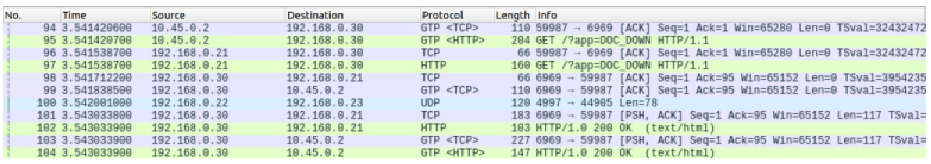}
  \caption{Proper UE Application Request Routing}
  \label{mfig:mfig6}
\end{subfigure} \\
\caption{E2E System Validation via Network Monitoring}
\label{mfig:mfig}
\end{figure*}

\begin{table*}[!htbp]
\label{entity}
\centering
\caption{Sample E2E Network Topology}
\begin{tabular}{|c|c|}
\hline
\textbf{IP}  & \textbf{Network Entity} \\ \hline
192.168.0.12 & NRF                     \\
192.168.0.13 & AMF                     \\
192.168.0.14 & SMF                     \\
192.168.0.15 & AUSF                    \\
192.168.0.16 & UDM                     \\
192.168.0.17 & UDR                     \\
192.168.0.18 & PCF                     \\
192.168.0.19 & NSSF                    \\
192.168.0.20 & BSF                     \\
192.168.0.21 & UPF1                     \\
192.168.0.22 & gNB                     \\
192.168.0.30 & UE                      \\
192.168.0.32 & UPF2                      \\ \hline

\end{tabular}
\label{entity}
\end{table*}

The following presents a data-based validation of full E2E network system operations through network monitoring and packet capture mainly utilised by the NWDAF. Figure \ref{mfig:mfig} highlights sample communication sequences during the instantiation and normal service operations of the realized E2E 5G+ network. The purpose of each subfigure is to demonstrate proper functionality of the core and access networks in addition to user application requests and responses. To supplement the information in Figure \ref{mfig:mfig}, Table \ref{entity} lists each entity in the E2E 5G+ network, from core NFs to gNBs and UE, and their respective IP in the sample private network. For this example, two UPF instances are instantiated and managed to demonstrate communications with multiple instances of the same NF.\par

Figure \ref{mfig:mfig1} shows a sample service operation named \textit{NF Status} performed by the NRF which is responsible for monitoring status information for NFs in the core network. The NRF communicates with the AUSF and AMF in this sequence, and the HTTP POST request represents the updated information that the NRF is tracking. The importance of this sequence is to confirm HTTP/2 interactions over the SBI in the 5GC which is typically on port 7777 by default.\par

Figure \ref{mfig:mfig2} validates the Packet Forwarding Control Protocol (PFCP) association between the SMF and UPF. When both multiple SMF and UPF instances are instantiated, a one-to-one Association Setup Request/Response must exist for all SMF and UPF instances in the private network. Two requests from two UPF instances in the sample network are present in this valid sequence.\par

Figure \ref{mfig:mfig3} demonstrates a similar NF association to the previously discussed sequence before. It highlights the connection of the core and access network through the AMF and gNB, respectively. When the gNB registers with the core network as an access point, the gNB and AMF communicate through Next Generation Application Protocol (NGAP) Setup and through Stream Control Transmission Protocol (SCTP). After this sequence is seen through network monitoring, the gNB is successfully registered with the AMF and the core network.\par

Figure \ref{mfig:mfig4} displays the full UE Registration sequence as it communicates with the core network through the active gNB. It is important to note that the source and destination IPs indicate the UE is communicating to the AMF for access management through the gNB or access network. In addition to NGAP, the UE Registration request uses the Non-Access-Stratum-5G-System (NAS-5GS) protocol for maintaining the uplink/downlink connection in radio communication and for establishing a Protocol Data Unit (PDU) session with the gNB and AMF. This sequence is vital to ensuring proper UE registration before testing user applications through the E2E network.\par

Figure \ref{mfig:mfig5} showcases a sample web application request issued by a user device through the E2E network. It is paramount to note that the IP, \textit{10.45.0.2}, is not present in Table \ref{entity} as the UE is not part of the E2E system's private network; UE addresses are assigned for the purpose of communicating with the network externally through the UPF as is the case for real wireless infrastructure. The HTTP GET request performed by the user device is encapsulated into GTP as the UPF tunnels communications between the gNB (from the UE) to the destination network, or web application server. Copies of the same GTP packets demonstrate how the UPF is responsible for proper UE application request routing to the destination address, and the original source address of the request will be the UE IP, \textit{10.45.0.2}.

\subsection{Network Analytics and Intelligence}

The NWDAF is introduced to discuss network data analytics prototyping within the emulated 5G Core. The NWDAF, as per the 3GPP standard \cite{3gpp.23.288}, is responsible for data analytics and network learning, and represents operator-managed network analytics as a logical function. It also provides slice-specific network data analytics to any given NF or consumer. As well, the NWDAF provides network analytics information to NFs on a network slice instance level and it is not required to be aware of the current subscribers using the slice \cite{sattar2019optimal}. In 5GC, the PCF and the NSSF are consumers of the network analytics that the NWDAF provides in the form of events subscription. The PCF uses these events for steering traffic policies and formulating policy decisions, and the NSSF uses load-level information provided by the NWDAF for guiding slice selection \cite{rost2016mobile}.

As of September 2022, the 3GPP have outlined a detailed standard and copious specifications for network data analytics services and the NWDAF \cite{3gpp.23.288}. The NWDAF leverages ML solutions through the Analytics Logical Function (AnLF) and the Model Training Logical Function (MTLF). The AnLF performs inferencing, or predictions based on analytics consumers' requests, on derived analytics information and statistics. The MTLF trains ML models with analytics information (either statistical information of historic events or future predictive information).

The NWDAF is intended for continuous monitoring of every NF, network slice, and UE device within the 5GC, and industrial NWDAF implementations provide closed-loop automation for third-party NFs in the 5GC. It is the responsibility of the creators of the NWDAF, who design and synthesize analytics services and techniques, to formulate Key Performance Indicators (KPIs) to measure network performance, which is discussed in the next sub-section. The real-time KPIs can be used for prescriptive analytics to automate network issue resolution, as labels/outputs for AI/ML problems, by recommending decision-making to improve performance, while predictive analytics can be used to predict future network issues. Predictive analytics also involves anomaly detection, which is used to automate mitigation within the 5G network \cite{radcom_2021}. 

\subsection{KPI Measurement and Evaluation}

KPIs are synthesized from collected metrics in the 5G network using such E2E testbeds. They are used for evaluating certain network performance cases and testbeds can use them in monitoring for real-time data analytics. Monitoring within the 5G Core has been worked on and assessed using the aforementioned functional NWDAF implementation in this paper, which employs packet capturing and SBI monitoring techniques to collect data on 5G microservices and service operations within the emulated 5GC. The high-computational server for the core network, referenced in Figure \ref{systemproto}, running Hyper-V as its hypervisor hosts multiple Ubuntu virtual machines with the emulated 5GC NFs, gNBs, and UEs. The \textit{MongoDB} instance and the web server also reside within the private network that the 5GC is hosted on.

Network monitoring techniques leverage the understanding of the service-based architecture (SBA) framework that the 5G Core uses. In accordance with the 3GPP standard, the architectural elements of the 5G SBA framework are defined in terms of NFs, rather than by traditional network entities. All the network functions communicate with one another via common interfaces or reference points. A network function that can monitor these reference points through a common interface can provide services to other authorized network functions as necessary \cite{rost2016mobile}, including associated microservices and responsibilities with the UPF (which handles user data), the external Data Network (DN), as well as other NFs (AMF, SMF, etc.) \cite{nokia}.

The importance of an analytics engine drives pre-processing for the NWDAF in order to generate model-ready datasets. Open-source distributed streaming systems, such as \textit{Apache} \textit{Kafka} \cite{Apache}, provide certain functionalities useful for the implementation to use in monitoring and pipelining the captured network data. \textit{Kafka} \textit{Connect} is used to stream the data to a \textit{MongoDB} instance, through source and sink connectors (source and destination). This technique is utilized to perform historical aggregate data analytics, as well as current state monitoring for future policy decision-making in the 5G Core. The 5G Core emulation operates in the private network without Network Address Translation (NAT), so port forwarding rules are configured for UE device traffic to be encapsulated in GTP when communicating with the UPF. The UPF can, thereby, route web application connectivity to UE devices for the purpose of running sample applications on UE devices.

Using data collection and data siphoning techniques is essential in gathering data into a network analytics repository, as the NWDAF implementation does. Network monitoring through the NWDAF can automatically update current packet capture records (using \textit{Kafka} Change Data Capture operations) for 5GC operations. When multiple UE devices connect to sample applications through the 5GC, newly generated packets are processed and transformed into schema-validated NWDAF events and can be used by any algorithm or other streaming transformation to pre-process the data ready to be input into an ML model. The E2E 5G+ system, alongside the NWDAF, demonstrates the capability of closed-loop automation in its entirety, in order to maximize the potential of the NWDAF and its impact on operation-specific decisions and network maintenance in an E2E 5G+ system.

\begin{table*}[!htbp]

\centering
\caption{Total Number of Packets Exchanged Within E2E 5G+ System During Multiple UE Requests to Single Web Application Server and when UE is Idle}
\begin{tabular}{|c|c|c|c|}
\hline
\multicolumn{1}{|l|}{\textbf{Network Entity}} & \textbf{No. of Packets (Idle UE)} & \textbf{No. of Packets (1 Requesting UE)} &  \textbf{No. of Packets (500 Requesting UEs)}\\ \hline
UE          & 0       & 5  & 2502   \\
gNB         & 19      & 25  & 2525 \\
UPF         & 2       & 9  & 2504 \\
Application Server & N/A       & 10  & 5004 \\
NRF         & 18       & 22 & 22 \\
AMF         & 10       & 11  & 12 \\
SMF         & 14       & 18  & 15 \\
AUSF        & 3       & 3  & 3  \\
UDM         & 3       & 6  & 6  \\
UDR         & 3       & 5  & 3   \\
PCF         & 3       & 5  & 3   \\
NSSF        & 3       & 3  & 3   \\ \hline

\end{tabular}
\label{packets}
\end{table*}

For the purpose of illustrating the necessary exchanges or number of messages passed through the entire E2E 5G+ system during usage, Table \ref{packets} shows the number of packets sent between 5G Core NFs and between the UE and application server. Specifically, UEs, having already been registered to the network, request a 476.23 KB text document from a web application server within a 10-second time window. It considers that the UE devices send packets for the web request, as observed within packet capture, by GTP tunneling through the UPF as it encapsulates the packets, adds more header content to the packets as they travel, and introduces new exchanges between the gNB and the UPF. The "Idle UE" case in Table \ref{packets} refers to the fact that the UEs are already registered, but still send messages through the RAN, which can be heartbeat messages for example. This assessment is paramount to understanding the synthesis of KPIs from throughput metrics and measuring the network response (the 5GC response, specifically) when UEs are using their respective applications. As well, we illustrate the scalability of the network response for each NF in the 5G Core, especially between 1 and 500 requesting UEs, for the purpose of understanding the level of resources required for a group of users to access their applications within the network. The number of packets for some NFs is unaffected by the number of users and this is because no new service operations are required for application requests. The common value of 3 indicates heartbeat responses to the NRF which monitors NF statuses in the network. The number of packets for the UE, gNB, and UPF are not equal and this is due to the retransmission of data between virtual machines during data collection.

\begin{figure*}[!htbp]
\begin{subfigure}{2\columnwidth}
  \centering
  \includegraphics[width=\columnwidth]{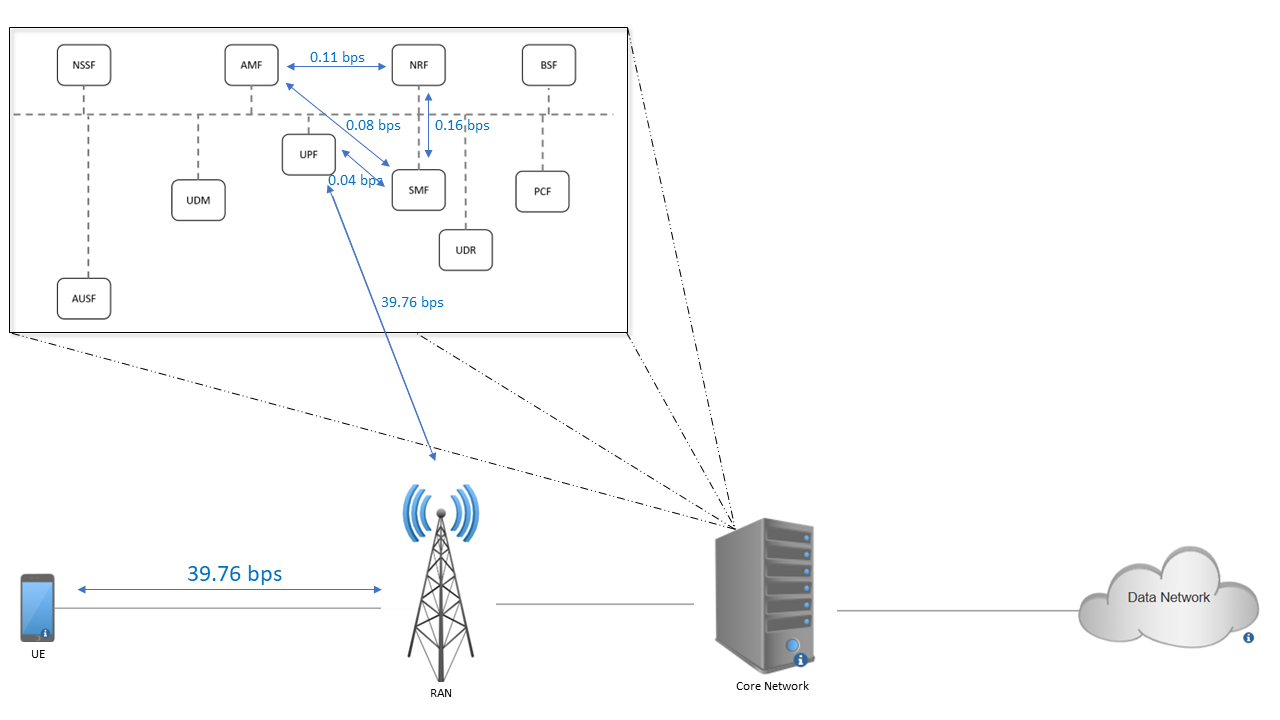}
  \caption{Average Network Throughput During 1 UE Web Application Request}
  \label{tfig:tfig1}
\end{subfigure}\\ \\
\begin{subfigure}{2\columnwidth}
  \centering
  \includegraphics[width=\columnwidth]{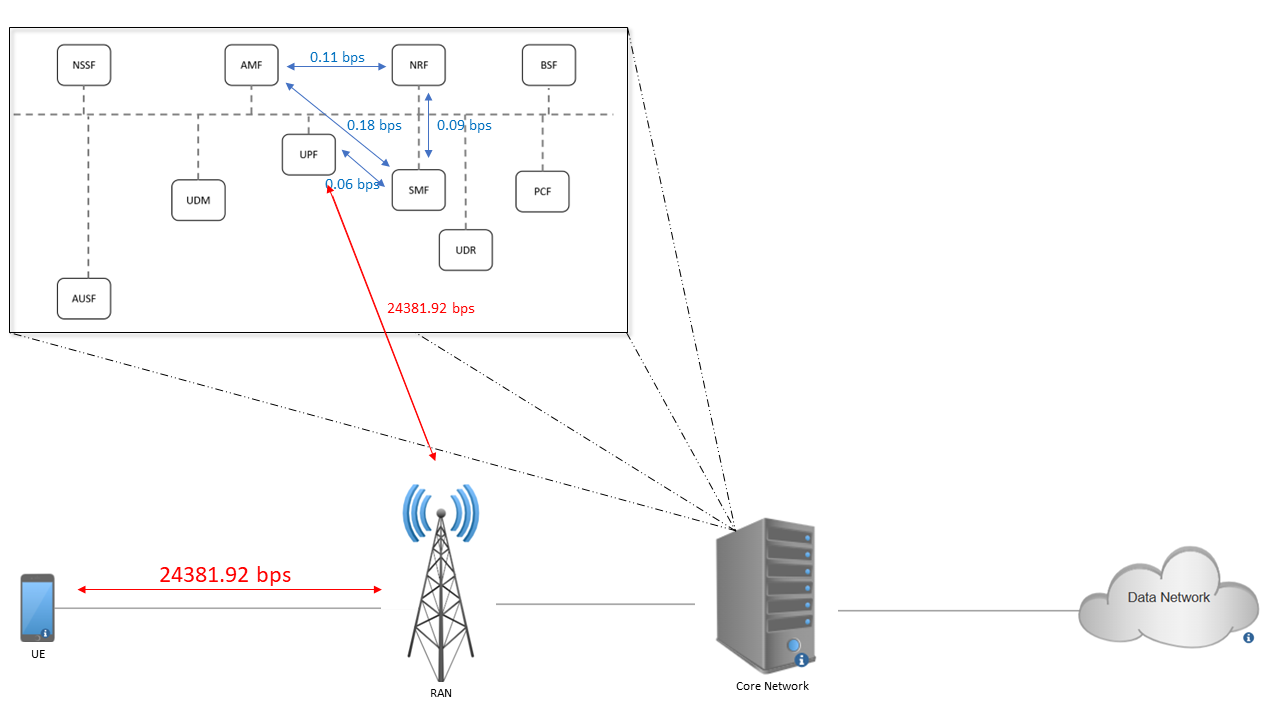}
  \caption{Average Network Throughput During 500 UE Web Application Requests}
  \label{tfig:tfig2}
\end{subfigure} \\ \\
\caption{Network Throughput Analysis For Users Accessing Applications Through E2E Network}
\label{tfig:tfig}
\end{figure*}

Figure \ref{tfig:tfig} illustrates a network analysis of E2E throughput within the E2E 5G+ system during UE application use. Specifically, insights into 5G Core control plane throughput are correlated with UE web application requests for analyzing the system response to handling application requests. One UE device issues an HTTP GET application request, in Figure \ref{tfig:tfig1}, through the E2E 5G+ system and the bi-directional throughput values are shown in blue lines. Between the UE and the gNB, the network throughput value is expected as the UE has already registered in the network and only issues an application request. The cross-section in the top-left of Figure \ref{tfig:tfig1} shows the 5G Core as the core network and the throughput values between NFs is expected: the associated packets consist of heartbeat messages and other periodic SBI interactions that are shown through packet capture and network monitoring. Figure \ref{tfig:tfig2} shows the throughput values for 500 UE devices issuing a web application request at once. The bi-directional throughput values for the UE and gNB are highlighted in red to demonstrate the contrast between a concentrated load of requests to the core network and the interactions within the 5G Core itself. The core network response to a large number of UE devices is very important for identifying performance bottlenecks in handling application requests and maintaining a stable network state. In terms of network management, orchestrating and scaling core NFs is vital to ensuring service operations remain stable and that the E2E system does not hinder the UE from accessing applications in a timely manner.

\subsection{Network Security and Data Privacy}

Core networks that employ open-source solutions, such as \textit{Open5GS}, require TLS v1.3 encryption over the HTTP/2 protocol for NF-NF communications in the 5G Core \cite{kone2022network}. As such, decryption of data analytics processes, in the forms of packet capturing and SBI monitoring, is assessed as a functional requirement of an E2E 5G+ system. 

TLS v1.3 is already an important obstacle in monitoring and inspecting network traffic for enterprises. Such monitoring use cases include investigating data breaches, malware, malicious activity, and performance issues within the network. A solution implemented by the realized E2E 5G+ system uses decryption libraries provided by \textit{Nubeva} software. \textit{Nubeva} employs Session-Key Intercept (SKI) to allow sensors embedded within the network to discover and ascertain TLS client/server session secrets without the need for certificates or session replay. These sensors extract the TLS keys and send them to their \textit{FastKey} server, which is responsible for sending those keys to the node containing traffic monitoring for the purpose of decrypting the network traffic from the network tap or packet capturing node. In other words, the traffic monitoring node that comprises the NWDAF is responsible for running the SKI decryptor necessary to decrypt the encrypted data between 5G Core NFs. \textit{Nubeva} was chosen for its high-performance decryption library that is paramount to high decryption throughput for truly real-time data analytics provided by the NWDAF \cite{skis}.

\section{Features of the Current E2E 5G Network System}

The following section presents the features of the current E2E 5G network system as high-level areas of application in wireless networking. Each feature represents how the E2E system is designed to enhance, test, and improve these features for researchers and business consumers utilizing the complete modular system. The design considerations, especially the modularity, of the testbed were made to enable the following set of features. The testbed, in its current state, can support each of the mentioned features and will continue to do so in all subsequent iterations and updates. 

\subsection{Cybersecurity Testing \& Application}

By 2025, the global annual cost of cybercrime is estimated to reach 10.5 trillion USD \cite{Morgan_2020}. As networks become increasingly distributed, the increased number of points of presence also increases the number of targetable interfaces that can be exploited. The impact of cybercrimes on a person, institution, or organization is profound. Aside from the financial cost of the attack, there is also the cost of a damaged reputation, the cost of a lack of trust going forward, and even a physical cost (i.e., failure of critical infrastructure leading to a risk of or loss of life). As the cyber threat landscape transforms and evolves, so too must the mitigation landscape in order to ensure the safety and privacy of all network users and service providers.\par
The emerging and prevalent cyberattack incidents in current networks include ransomware, malware, phishing, data breaches, and denial-of-service attacks. Each of these attacks can target various network entities and has a unique infiltration point. \par
\begin{itemize}
\item Ransomware attacks gain access to a system and then lock it and encrypt its contents. System access is only restored after a payment to the malicious agent is made. Globally, the frequency and severity of ransomware attacks are on the rise, with a case that included loss of life when critical medical infrastructure was targeted \cite{Stouffer_2022}.  
\item Malware attacks occur when a piece of software gains access to a system (usually through accidental installation) without the user’s knowledge. This software then proceeds to access, destroy, or compromise system information and elements. Malware can have many shapes and can target even the largest of entities, as seen by the 2021 Microsoft Exchange Server attack \cite{Microsoft_2021} or the 2022 NVIDIA attack \cite{NVIDIA_2022}.
\item Phishing attacks occur when malicious entities impersonate trustworthy sources to access private or secure information. Common phishing scams target social security numbers, online bank account information, credit card numbers, and account passwords. 
\item Data breaches occur when malicious users gain access to information without proper authorization. As global systems become more and more data-centric with the rise of AI, these attacks will become increasingly common and more severe. In 2023, T-Mobile, one of the USA’s network service providers, faced a data breach event which impacted an estimated 37 million user accounts \cite{T-Mobile_2023}. 
\item Denial-of-Service (DoS) and Distributed Denial-of-Service (DDoS) attacks are some of the most notorious and damaging cyberattacks. In these attacks, a malicious user attempts to render a network or service inaccessible by flooding it with requests and overloading the system’s request handling capacity. In early 2020, Amazon Web Services (AWS) was hit with the largest DDoS attack ever recorded, with a peak request rate equaling 2.3 Tbps \cite{AWS_2020}. 
\end{itemize}
From the mentioned attack types, it is clear that vulnerabilities in any aspect of the network will be exploited by malicious users. These exploits can have devastating impacts depending on their size and nature. As Industry 4.0 develops, and IoT proliferation continues, there is a credible fear that compromising these devices will help malicious actors infiltrate various users and organizations. To date, cyberattacks have targeted communications, healthcare, finance, government, education, energy, and transportation entities and institutions. It is evident that cybersecurity has a critical role to play in the protection of the privacy and security of future networks and users. Given the increasing smart connectivity and autonomous transportation on the horizon, this protection now includes the safeguarding of human life and is of paramount importance. \par
To this end, the modularity of the testbed supports the entire cyber threat deterrence pipeline. Each network element, entity, technology, and region can be rigorously tested to identify vulnerabilities. Frameworks to actively protect against cyber threats and safeguard the system can be deployed and tested. Using these deployed frameworks, cybersecurity threats can be identified, and an appropriate response can be taken to mitigate the issue. Policies related to recovery in the case of impact can also be implemented and enacted. Two critical features supporting various stages of this pipeline include Penetration Testing (Pen Testing) and Intrusion Detection Systems (IDSs).\par
\subsubsection{Penetration Testing}
Pen testing is a process by which a simulated cyberattack is conducted to determine the vulnerabilities of a system or entity. The basic phases of a pen test include planning and reconnaissance, scanning, gaining access, maintaining access, and maintaining anonymity. There are various pen testing categories, including external testing, internal testing, anonymized testing, double-anonymized testing, and targeted testing \cite{Cloudflare}. The scope of a pen test can include network infrastructure, services and applications, wireless networks, social engineering, and physical systems \cite{Cisco_2023}. The result of all pen tests is the analysis and mitigation of system vulnerabilities and the overall improvement in cybersecurity threat mitigation readiness. \par
\subsubsection{Intrusion Detection Systems}
The IDS is an ML-based anomaly detection framework that monitors network traffic and identifies anomalous behaviour. Due to the disproportionate amount of normal traffic compared to abnormal traffic, IDSs often build a model of normal or expected behaviour and use that as a basis to determine what constitutes abnormal traffic. This approach is more intuitive than directly modelling abnormal traffic since the threat landscape is constantly changing. By characterizing normal behaviour, IDSs are more likely to be robust to future threats without explicitly having seen them during training. When an IDS encounters abnormal traffic, its first step is to block the traffic and raise an alarm. There are various types of IDSs, including core-based, IoT-based, and cloud-based, and they can, therefore, be deployed throughout the network on various elements, systems, and regions.\par

\subsection{Network Data Analytics Towards Automated Network Intelligence}

The profound impact ML/AI has had on the world in recent years is undisputable. Global research institutions, industrial sectors, and governments are devoting time and resources to the development and adoption of intelligence methods. In the telecommunications sector specifically, the AI market size is expected to reach USD 38.8 Billion by 2031 \cite{Deotale_2023}. AI has the power to revolutionize network operations through boosted performance and effectiveness, enhanced sustainability and energy efficiency, improved security and trustworthiness, as well as the creation of new business verticals. Some of the major challenges that relate to ML/AI adoption in networking relate to data availability and data infrastructure planning. The development of this testbed has addressed both of these challenges by being Machine Learning Operations (MLOps) focused throughout the entire architecture and design process. \par
MLOps defines the practices and considerations that go into the deployment of ML solutions. There are various stages that go into the MLOps pipeline, including data ingestion and preparation, model training and tuning, and the deployment and continual monitoring of the system. The development and integration of the NWDAF, a critical part of the testbed, is evidence of the successful consideration of MLOps during the architectural design process. With the streamlined data pipeline, from generation to consumption, and the various model deployment and monitoring frameworks, the testbed is positioned to upgrade its AI/ML deployments as we transition into the beyond 5G era. \par
The development and integration of the NWDAF, as well as the infrastructure required to support the entire MLOps process, are the first steps towards Zero-Touch Network Service Management (ZSM). The ZSM paradigm is an automation process that defines self-configurable, self-healing, and self-optimizing networks. This paradigm is built upon sensing, monitoring, and predicting through analytics and intelligence. Through ZSM, the prospect of proactive network management becomes a tangible possibility. By sensing the network and predicting its future state, proactive management decisions can be enacted to mitigate unfavourable forecasted conditions. The idea of ZSM will play a defining role in future networking generations, building upon the foundations laid out in this testbed and, more broadly, in 5G networks.\par

\subsection{User Application Diversity}

One of the defining characteristics of 5G networks is the number of new and revolutionary user applications they can support. In order to support these emerging and constantly evolving technologies, QoS requirements must be ensured, and Service Level Agreements (SLAs) must be upheld. To this end, as current user applications evolve and new ones are developed, the network requirements needed to support them must be accounted for and the appropriate resources provisioned. The architecture of the developed testbed was designed to meet the requirements of current user applications and offer the flexibility to support future applications. The modularity and interoperability of the elements in the testbed allow for modular updates and modifications to network elements as required to support new features. Additionally, extensive slice management capabilities that address all aspects of the slice lifecycle ensure that dedicated resources are available to critical and high-performance applications, logical isolation is ensured for secure entities, and resilience schemes are in place to mitigate the effect of an unforeseen fault or failure and ensure service continuity.

\section{Future Outlook for E2E Network Testbeds: The Transition to Beyond 5G}

\begin{figure*}[!htbp]
\centerline{\includegraphics[width=1.2\textwidth]{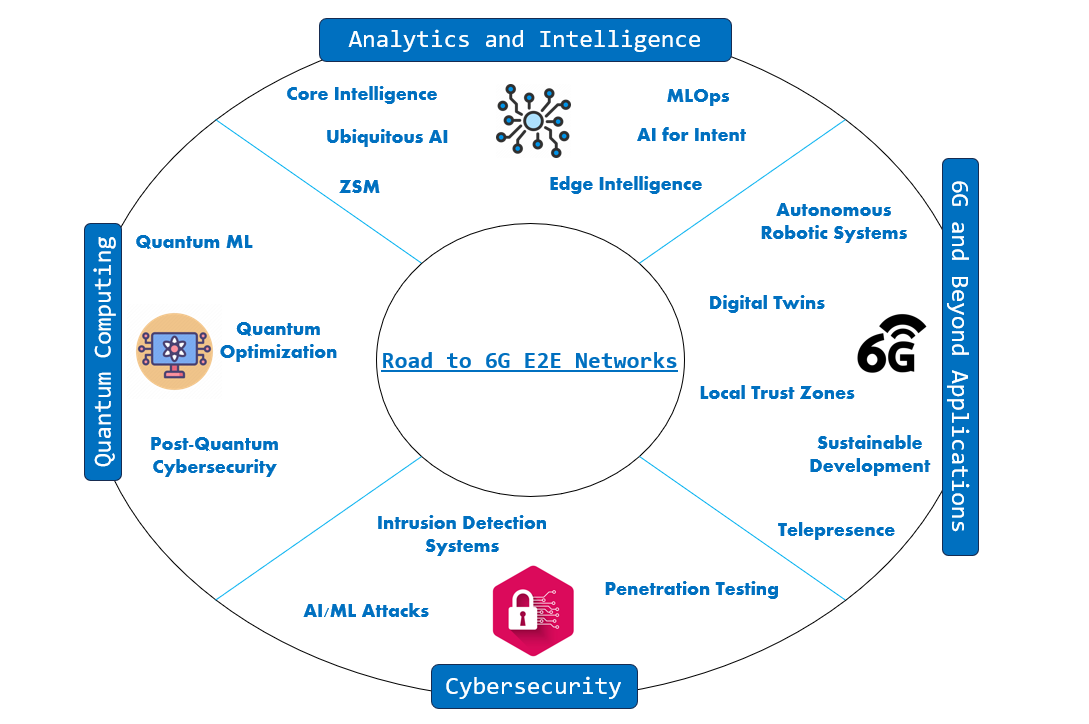}}
\caption{Road to 6G E2E Systems Mindmap}
\label{roaddia}
\end{figure*}

The following section presents a future outlook for the development of network management platforms using E2E networks as a tool for MNOs and CSPs alike to support emerging technologies for enabling full network automation. Specifically, the limitations of the currently implemented E2E 5G+ system are discussed and assessed towards the transition to novel applications and use cases in 6G and future networks. Figure \ref{roaddia} illustrates a mindmap detailing the important avenues of research and development on the road to 6G E2E systems.

\subsection{Limitations of Current E2E 5G+ System}

The following section considers the limitations of the realized E2E 5G+ system with respect to future modifications for enhancement and improving network analytics functionalities for the purpose of exposing more of the 5G network to MNOs and business consumers of 5G solutions. 

The limitations of the 5G+ system consider the effectiveness of the NWDAF implementation in predicting and prescribing analytics, as well as the degree to which full 5GC and RAN functionalities are realized within any conceived network topology. One assessed limitation in literature that has become apparent in synthesizing the system environment is the lack of "organic adaptation". This term is used to explain the necessity of merging and regrouping 5G NF functionalities with respect to specific use case needs \cite{corici2022organic}. This is important for both NWDAF development and future considerations for the NWDAF in 6G networks. The former is evident through the introduction of the Application Function (AF) which is not integrated with any of the open-source 5GC solutions tested and discussed within this paper; the AF is intended to introduce another generic communication interface with the other core NFs to assess application requirements for using existing services or introducing new ones. The latter stresses the important changes that an NWDAF implementation must realize when considering the stringent requirements of emerging 6G networks.

Another limitation of the realized E2E 5G+ system is the lack of a dynamic slicing manager in the ETSI NFV-MANO framework, which is essential to network slice lifecycle management. The automation of slice creation, maintenance, and deletion are very important to a closed-loop testbed. A dynamic slicing manager can support provisioning for concurrent applications while ensuring the aggregated requirements are feasible at all stages. When considering the manager's effectiveness in utilizing available resources in a reasonable fashion, there arises a new limitation in balancing system performance, experienced QoE, and operating costs with existing and new network slices \cite{kapassa2018dynamic}.

\subsection{Quantum Systems}
Quantum computing is a revolutionary technology that has garnered significant attention since its inception. With the promise of extremely enhanced computational capabilities, these computers are envisioned to help solve some of the most complex computational problems. Despite not being readily available, the potential of these computers has already impacted several industries and is actively being considered for future financial services. Additionally, due to the increased computing capabilities, various machine learning and complex optimization tasks that face execution feasibility challenges on classical systems are looking towards quantum for efficiencies and improvements. One of the key milestones for quantum adoption is the practical demonstration of quantum advantage over classical systems beyond a theoretical perspective. Despite the numerous opportunities for innovation using this technology, there are several challenges, especially related to cybersecurity, that are causing great global concern and prompting a change in our current practices. \par
\subsubsection{Post-Quantum Cybersecurity}
The main source of concern for quantum computing resources is its effect on current cybersecurity systems. The ability to quickly and efficiently factor large numbers has the potential to render all current encryption systems obsolete. To this end, global standardization agencies have already pushed towards the development of post-quantum cybersecurity measures, including encryption, protocols, and frameworks. As this technology matures, its integration into future networks will be of paramount importance for the safeguarding of private data and the security of network systems. As the methods of malicious agents attempting to breach the network and compromise its various entities become more and more sophisticated, so must the various cybersecurity defence and deterrence mechanisms and frameworks. \par
\subsubsection{Quantum Optimization}
As networks become more distributed and complex, their management and orchestration become increasingly important for performance and efficiency requirements. The prospect of ZSM has led service providers to explore options for real-time optimization and re-configuration. Quantum-based optimization has emerged as a key enabler of real-time optimization, especially for critical applications in future networks \cite{duong2022quantum}. Before widescale adoption, several challenges related to quantum optimization must be solved. Firstly, the availability and cost of powerful quantum resources is a major deterrent to adoption. Additionally, the current storage limits of quantum computers limit the complexity of the problem that can be solved \cite{wang2023opportunities}. Finally, expressing a problem existing in a non-quantum space as a quantum problem requires efficient encoding and embedding schemes. The field of quantum optimization, specifically its integration in future networks, is a rapidly developing field with incredible promise and potential for attaining real-time optimization.
\subsubsection{Quantum Machine Learning}
One of the most discussed applications of quantum computing is its integration with machine learning methods and the creation of Quantum Machine Learning (QML). QML is projected to be the next frontier in intelligence research. The use of a higher dimensional dataspace (\textit{i.e.,} Hilbert space) to represent intricate data enables the extraction of complex and previously undiscovered features and relationships that would go unnoticed in classical dataspaces. Furthermore, the use of quantum-native algorithms to improve the efficiency of the processes and inner workings of machine learning models creates a plethora of possibilities. As with any technology in its infancy, significant challenges, including quantum-native data availability, quantum black-box interpretability, and quantum resource big data processing and storage, must be solved before it inspires trust and is ready for integration with future networks and networking systems.
\subsection{Ubiquitous AI}
Intelligence is one of the most talked about aspects of current and future networks. In 5G, network intelligence is limited to network functions (i.e., NWDAF, MDAF, etc.). While the creation and standardization of the functions in the 5G core architecture was a pivotal moment for intelligence integration in modern networking systems, it is only the proverbial ‘tip of the iceberg’ in terms of unlocking its true potential. Future networks will have a much more profound integration of intelligence extending far beyond a set of network functions deployed in the core. The ubiquitous deployment of intelligent entities woven into the fabric of the network will push the boundaries of innovation and unlock a myriad of possibilities. 
\subsubsection{AI at the Core}
Future core networks are envisioned to be AI-native, meaning that they are designed to include widespread AI deployments with the appropriate data pipelines and frameworks throughout. This approach is different from previous networking generations as AI will be a foundational element of the network rather than an extrinsic addition. The pervasive nature of future AI in the core will enable improved performance with a lower complexity \cite{nokia}. Another key benefit of an AI-native core network is the ability to be considered organic and perform morphology evolution, a process which can lead to new network components, the consolidation of existing network components, the removal of unnecessary network components, and the redefinition of network component interactions \cite{corici2022organic, lu2023architecture}. Essentially, an AI-native network core will resemble a biological organism that reacts to environmental stimuli and evolves over time to attain benefit or efficiency.
\subsubsection{AI at the Edge}
Multi-Access Edge Computing (MEC) was one of the fundamental architectural components of 5G networks and enabled a plethora of use cases relating to Industry 4.0 and the vision of smart cities. In 6G, this technology will host critical deployments of intelligent entities. These intelligent entities will be responsible for the collection of environmental data, computation leading to cognition, and management and orchestration \cite{wang2019edge}. The ability to deploy AI on lightweight points of presence will allow for true pervasive intelligence in all regions and will enable revolutionary use cases related to transportation, agriculture, and environment \cite{yang2020artificial, manias2021making}. Some of the key considerations for an AI-native network edge include resource-efficient training and inference, edge-enhanced ML models and architectures, scalable edge deployments, and enhanced security protocols \cite{letaief2021edge}.
\subsubsection{AI for Decision Making}
One of the greatest prospects of AI is decision-making. In order for an intelligent entity to be capable and trustworthy of making decisions, machine reasoning must be attained. In its essence, machine reasoning is the use of logic to make decisions in a process that mimics the human decision-making process. In order to attain machine reasoning, an entity must be able to process incoming information and self-learn from it. However, the ability of an AI agent to self-learn and display evidence of logic is not enough to make it trustworthy for higher-order decision-making. The trustworthiness of AI agents will be based on their explainability and interpretability. In order for a decision to be trusted, the agent must explain the factors that led to that decision. Additionally, the factors that lead to a decision must be interpretable to ensure that they are absent of implicit bias. 
\subsubsection{AI for Intent}
Recent developments in AI, specifically Large Language Models (LLMs), have made the prospect of true E2E automation attainable through intent extraction. By leveraging these powerful models to extract user intent, the process of transforming this intent into an action is streamlined. LLMs will interface directly with customers and autonomously complete the request or forward the request to an appropriate entity. A prime example of this would be if a customer requests a service change or a network reconfiguration; the LLM would extract the intent and relevant information and make the required changes. Another customer-facing interaction for LLMs would be live troubleshooting and customer support. By deploying these LLMs, performance metrics such as mean response time and mean time to request completion can be improved, thereby directly impacting customer satisfaction and operational efficiency. The use of LLMs for intent is not only restricted to customers. Network engineers can leverage these models when reconfiguring networks or scaling resources; the engineer would converse with the LLM, their intent would be extracted, the required action would be taken, and the results of the action would be relayed back to the engineer for review. The integration of LLMs into future networks is currently in its infancy; however, given their success in various industries since their inception, these models have a wealth of untapped potential.
\subsection{Next-Generation Applications}
The evolution of networking generations is often motivated by the changing application and use case landscape. To this end, a set of use cases pushing the limits of the envisioned performance of 5G networks and catalyzing the need for 6G development are summarized below. The provisions made during the testbed development allow for modifications and upgrades to support these future services and applications. 
\subsubsection{Telepresence}
One of the most lucrative visions of the future is the idea of telepresence. Telepresence is defined as a remote multisensory interaction with a physical or virtual system \cite{Rugeland_2021}. Currently, the Metaverse is an initial development of telepresence as it is an online virtual reality world that has the capability to interact with the real world \cite{chang20226g}. In the future, alongside the rapid development and expansion of the Metaverse, holographic telepresence will emerge, where remote users are fully rendered in the real world. The future of telepresence is contingent upon the developments in virtual, augmented, and mixed reality, the network capabilities, and the effective modelling of the physical world in the virtual domain. Telepresence, as a whole, is a data-intensive process and requires real-time sensing, transmission, and rendering. To this end, future networks will require enhanced reliability, faster transmission rates, and lower latency beyond what 5G has to offer to provide a fully immersive and uninterrupted experience \cite{tang2022roadmap,peng20226g}. 
\subsubsection{Autonomous Robotic Systems}
The envisioned robotic systems of the future extend far beyond the current implementation of robots, which is predominantly restricted to manufacturing settings and processes, to a much more profound integration into future societies. The majority of current robotic systems are designed to follow a specific set of commands, are explicitly controlled by an agent, and are isolated, or their interactions are limited to other robots. In future networking generations, robots will have a much more profound integration in our societies with the emergence of collaborative robots, known as cobots \cite{Fersman_2020}. The distinguishing feature of cobots compared to regular robots is their ability to interact, collaborate, and work alongside humans \cite{han2022multi}. One of the main requirements of cobots is the efficient perception of the environment, which, along with accurate localization, will enable seamless operation \cite{trevlakis2023localization}. Future networks, through their enhanced reliability and QoS, will enable the widespread deployment and societal adoption of cobots.
\subsubsection{Digital Twin Architecture}
Digital twins have revolutionized the way we think about the physical world. The process of digital twinning creates a virtual representation of a physical system, entity, or process. This virtual representation is created using historical data and modelling. One of the key features of digital twins is the data and information exchange between the physical and virtual worlds \cite{guo2023five}. Currently, digital twinning is being considered in manufacturing processes; however, the future of twinning includes massive expansion to include extensive societal twinning, including utilities, food production, transportation, etc. As the digital twinning of the physical world ramps up, key considerations such as trustworthiness, real-time capabilities, scalability, and security are highlighted \cite{lin20236g,alkhateeb2023real}. Beyond 5G networks are required to ensure the reliable and efficient transfer of information between the physical and virtual worlds, the management of resources to scale the twin to match the changing physical system, the flexibility to enable twin-twin interoperability, and the security against cyberattacks that can poison the twin and the subsequent management decisions on the physical system.    
\subsubsection{Sustainable Development}
The prospect of sustainable development is a key demonstration of the power and the need for beyond 5G networks. By leveraging the ubiquitous coverage and reach of next-generation networks along with the enhanced sensing capabilities they provide, numerous sustainable development use cases arise. The advent of non-terrestrial networks also gives rise to wide-area sensor networks on a global scale. The enhanced reach of these sensor networks can help monitor environmental indicators, provide services, including healthcare, to underdeveloped and remote areas, as well as enhance global supply chain management \cite{Rugeland_2021}. Key aspects of Beyond 5G networks, including 3D coverage, widespread AI integration, improved energy efficiency, and massive device connection density \cite{vaezi2022cellular} are required to realize these future sustainable development use cases.
\subsubsection{Local Trust Zones}
The final emerging use case for beyond 5G networks is the development of local trust zones \cite{Rugeland_2021}. As the future envisioned use cases of future networks become more computationally intensive and require enhanced performance capabilities, the ability to deploy transient provisional networks becomes increasingly appealing. These temporary networks can meet the performance requirements of a specific application that would exceed the capabilities of a larger wide-area network. Furthermore, the development of these provisional networks also provides enhanced data protection for critical and sensitive information through logical isolation and boundaries \cite{Pietschmann_2020}. The use cases of these local trust zones regarding performance can include gaming and telepresence, whereas enhanced trust is beneficial to use cases such as healthcare. The ability to provide the necessary performance requirements, as well as data privacy (through containment and vulnerable surface reduction) at a local scale, is a powerful and critical enabler of future networking applications.

\section{Conclusion}

This paper presents the building process and foundation of an E2E 5G+ system, which can act as a network analytics testbed that enables the deployment, monitoring, and analysis of a realized E2E network. In particular, a detailed background on the motivation and architectures for 5G testbed designs and solutions was provided and their capabilities were assessed through the evaluation of industry-leading software implementations of the 5G Core, the RAN, and interacting UE devices. An elaborate comparison was conducted for a multitude of open-source 5G system solutions in emulation and practice. Notably, a fully realized, state-of-the-art, modular, E2E 5G+ network system was discussed and its usability for network analytics exposure was made evident through practical experimentation. A discussion into the current and future capabilities provided by the testbed and extending into future networking generations was provided. The E2E 5G+ system can pave the way towards fully automated network management, and enhance how network operators and business consumers leverage all the benefits of an evolved 5G+ network.

\bibliographystyle{IEEEtran}
\bibliography{samplev2}
\vskip -2\baselineskip plus -1fil
\begin{IEEEbiographynophoto}{Ali Chouman}
is a Ph.D. student in the Optimized Computing and Communications (OC2) Lab at Western University. He received his B.ESc degree in Electrical and Computer Engineering (2020) and his M.ESc degree in Electrical and Computer Engineering (2022) from Western University. His current research interests include artificial intelligence, machine learning, network predictive maintenance, along with predictive and prescriptive analytics design. 
\end{IEEEbiographynophoto}
\begin{IEEEbiographynophoto}{Dimitrios Michael Manias}
is a Postdoctoral Associate in the Optimized Computing and Communications (OC2) Lab at Western University. He received his B.ESc degree in Electrical and Computer Engineering (2018), his M.ESc degree in Electrical and Computer Engineering with a Collaborative Specialization in Artificial Intelligence (2019), and his Ph.D. in Electrical and Computer Engineering (2023) from Western University. He was awarded the 2023 Governor General of Canada Academic Gold Medal. His current research interests include AI and next-generation networks.
\end{IEEEbiographynophoto}
\begin{IEEEbiographynophoto}{Abdallah Shami}
is a Professor at the ECE department at Western University, Ontario, Canada. Dr. Shami is the Director of the Optimized Computing and Communications Laboratory at Western. He is currently an Associate Editor for IEEE Transactions on Mobile Computing, IEEE Network, and IEEE Communications Tutorials and Survey. Dr. Shami was the elected Chair of the IEEE Communications Society Technical Committee on Communications Software.
\end{IEEEbiographynophoto}

\end{document}